\documentclass[aps,print,twocolumn,showpacs]{revtex4}%
\usepackage{hyperref}
\usepackage{amsfonts}
\usepackage{amsmath}
\usepackage{amssymb}
\usepackage{graphicx}%
\begin{document}
\preprint{ }
\title[ ]{Rotational Symmetry of Classical Orbits, Arbitrary Quantization of Angular
Momentum and the Role of Gauge Field in Two-Dimensional Space\\ }
\author{Jun-Li Xin$^{1,2}$ and J.-Q. Liang$^{1}$}
\affiliation{$^{1}$Institute of Theoretical Physics and Department of Physics, Shanxi
University, Taiyuan, Shanxi 030006, China; $^{2}$Department of Physics and
Electronic Engineering, Yuncheng College, Yuncheng, Shanxi 044000, China}
\author{}
\affiliation{}
\author{}
\affiliation{}

\begin{abstract}
We study the quantum-classical correspondence in terms of coherent wave
functions of a charged particle in two-dimensional central-scalar-potentials
as well as the gauge field of a magnetic flux in the sense that the
probability clouds of wave functions are well localized on classical orbits.
For both closed and open classical orbits, the non-integer angular-momentum
quantization with the level-space of angular momentum being greater or less
than $\hbar$ is determined uniquely by the same rotational symmetry of
classical orbits and probability clouds of coherent wave functions, which is
not necessarily $2\pi$-periodic. The gauge potential of a magnetic flux
impenetrable to the particle cannot change the quantization rule but is able
to shift the spectrum of canonical angular momentum by a flux-dependent value,
which results in a common topological phase for all wave functions in the
given model. The quantum mechanical model of anyon proposed by Wilczek (Phys.
Rev. Lette. 48, 1144) becomes a special case of the arbitrary-quantization.

\end{abstract}
\volumeyear{ }
\volumenumber{ }
\issuenumber{ }
\eid{ }
\date{}
\received[Received text]{}

\revised[Revised text]{}

\accepted[Accepted text]{}

\published[Published text]{}

\startpage{1}
\endpage{ }
\maketitle

\section{\qquad introduction}

Recently, renewal of interest has been evoked to the fractional angular
momentum (FAM) in two-dimensional (2D) space in relation with the
correspondence between quantum mechanical wave-functions and classical
periodic-orbits\cite{Makowski1,Makowski2}. The FAM is only possible in 2D
multiply-connected-space, since in three- or higher-dimensional space, the
angular momentum being integer or half-integer is completely determined by the
commutation relation of angular-momentum operators. In 2D space, the angular
momentum operator has only one component, which does not give rise to a unique
determination of the angular-momentum eigenvalue, and the common belief of
integer-quantization is based on the $2\pi$-periodic boundary-condition which,
however, is not justified.

Wilczek in his pioneering work proposed for the first time a quantum
mechanical model consisting of a charged particle and magnetic flux in 2D
space to demonstrated the fractional eigenvalues of angular momentum known as
anyon \cite{Wilczek1,Wilczek2}. In the existence of gauge field, however, we
have both the kinetic angular momentum (KAM) and the canonical angular
momentum (CAM), which are different, because of the velocity-dependent forces
\cite{Silverman,Hamermesh,Liang1,Schulman}. It is true that "the generator of
rotation should be the CAM and is prescribed by Noether's theorem as a
conserved quantity"\cite{Jackiw}, which is gauge dependent and
integer-quantized \cite{Jackiw,Liang1,Schulman}. While the KAM is gauge
invariant dynamic quantity, generally fractional because of the gauge field
\cite{Jackiw,Liang1,Schulman}. Although the consistence of fractional CAM with
the Aharonov-Bohm (AB) phase-interference was shown long ago \cite{Liang1} it
remains a long standing open question whether or not the fractional CAM plays
a role in quantum physics. It is worthwhile to remark that the gauge potential
of an AB flux can only shift the angular momentum eigenvalues by a common
fractional number but cannot change the integer-quantization rule, namely the
eigenvalue-space of angular momentum is still $\hbar$ \cite{Liang1,Liang2}.

The existence of FAM in a wide class of 2D central potentials without gauge
field has been discussed more recently \cite{Makowski1,Makowski2} by the
localization of coherent wave functions on classical orbits, which imposes a
special boundary condition leading to the unusual angular phase of wave
functions. Following the interesting studies of Ref.
\cite{Makowski1,Makowski2}, which provide an exactly solvable model both
quantum mechanically and classically, we consider a charged particle in the
gauge field of magnetic flux-string perpendicular to the 2D plane with the
central scalar-potentials in addition and study the non-integer quantization
of angular momentum (NIQAM) in terms of quantum-classical correspondence.
Since the coherent state is constructed by the superposition of angular
momentum eigenstates and thus the probability clouds have to possess the same
rotational symmetry as that of classical orbits, which is not necessarily
$2\pi$-periodic. As a consequence the NIQAM with the level-space being greater
or less than $\hbar$ appears naturally along with the correspondence
principle. By the explicit calculation it is shown that only the shift of CAM
eigenvalues by the gauge field, which gives rise to a common topological phase
for all wave functions, is in agreement with the correspondence. The validity
of wave functions with a topological phase is further confirmed by the
expectation values of KAM operator, which coincide with the classical values
of KAM $\pounds ^{k}$. The long standing problem whether or not the FAM is
related to the CAM is resolved and the quantum mechanical model of anyon
proposed by Wilczek \cite{Wilczek1,Wilczek2} emerges as a special case of
$\mu=1$ with $2\pi$-rotational-symmetry of classical orbits.

More specifically, we refer here to the particular zero-energy states, which
can be obtained analytically for both classical- and quantum-solutions
\cite{Makowski3,Liang2,Makowski4}. On the other hand the zero-energy states
are of importance in various fields such as\textbf{ }the cold-atom
collisions\cite{Wang,Makowski5}, the construction of vortex lattices
\cite{Kobayashi}, and quantum cosmology \cite{Nowakowski6}.

\section{Classical orbits and rotational symmetry}

We consider a charged particle of charge $e$ and mass $m$ in a gauge field of
infinitely long magnetic flux-line of total flux $\Phi$ located at the origin
of 2D space and central scalar-potential of form \cite{Makowski1,Makowski2}
\begin{equation}
A_{0}(r)=-\frac{\gamma_{\upsilon}}{r^{2\mu+2}},\quad\gamma_{\upsilon}%
>0,\quad-\infty<\upsilon<\infty
\end{equation}

Where $r=\sqrt{x^{2}+y^{2}}$ and $\upsilon=2\mu+2$.

Outside the flux-line ( $r>0$), the Lorentz force on the charged particle is
always zero because of the vanishing magnetic field $\vec{B}=0$, while the
vector potential in the polar coordinate is seen to be
\[
\vec{A}=\frac{\Phi}{2\pi r}\overrightarrow{e}_{\varphi}%
\]
with $\overrightarrow{e}_{\varphi}$ being the unit-vector of angular
direction. In order to establish the quantum-classical correspondence, we
ought to evoke the canonical variables. In the polar coordinates $(r,\varphi
)$, the Lagrangian of the system is seen to be
\begin{equation}
L=\frac{1}{2}m[\dot{r}^{2}+(r\dot{\varphi})^{2}]-eA_{0}(r)+L_{WZ}%
\end{equation}
where $L_{WZ}=\alpha\hbar\dot{\varphi}$ is so-called the Wess-Zumino
topological interaction-term with the parameter $\alpha=\Phi/\Phi_{0}$ being
the dimensionless magnetic flux in the quantum-unit $\Phi_{0}=ch/e$. The
Wess-Zumino term does not affect the equation of motion but the initial value
of angular momentum. Canonical momentums corresponding to the coordinate
variables $r$, $\varphi$ are defined by%
\begin{align}
p_{r}  &  =\frac{\partial L}{\partial\dot{r}}=m\dot{r},\quad\\
\pounds ^{c}  &  =\frac{\partial L}{\partial\dot{\varphi}}=mr^{2}\dot{\varphi
}+\alpha\hbar
\end{align}
\ Here $\pounds ^{c}$ is CAM, while $\pounds ^{k}=mr^{2}\dot{\varphi}$ is the
KAM. Then the Hamiltonian is
\begin{equation}
H=\frac{p_{r}^{2}}{2m}+\frac{(\pounds ^{c}-\alpha\hbar)^{2}}{2mr^{2}}%
+eA_{0}(r),
\end{equation}
From canonical equations we find that both the CAM and KAM are conserved
quantity
\[
\frac{d\pounds ^{c}}{dt}=-\frac{\partial H}{\partial\varphi}=0,\quad
\pounds ^{k}=\pounds ^{c}-\alpha\hbar
\]
For the case of zero-energy and nonvanishing initial KAM $\pounds ^{k}%
=\pounds ^{c}-\alpha\hbar\neq0$ we have
\[
\frac{1}{2}m(\frac{\pounds ^{c}-\alpha\hbar}{mr^{2}})^{2}[(\frac{dr}{d\varphi
})^{2}+r^{2}]-e\frac{\gamma_{\upsilon}}{r^{\upsilon}}=0.
\]
We assume an initial-value that
\begin{equation}
\pounds ^{k}=\xi_{k}\hbar, \label{6}%
\end{equation}
where $\xi_{k}$ is an arbitrary dimensionless-quantity. Thus in the considered
case only the CAM is shifted by the flux. Introducing a dimensionless variable%
\begin{equation}
u=r/\tilde{a}_{c},\quad\tilde{a}_{c}=\frac{(2me\gamma_{\upsilon})^{1/2\mu}%
}{[\xi_{k}\hbar]^{1/\mu}}%
\end{equation}
we obtain the equation of particle trajectories such as
\cite{Makowski3,Daboul2}
\begin{equation}
(\frac{du}{d\varphi})^{2}+u^{2}=u^{4-\upsilon}=u^{2-2\mu} \label{8}%
\end{equation}
The general solution of Eq.(8) is given in Ref.\cite{Makowski3,Daboul2}
\begin{equation}
r^{\mu}=\tilde{a}_{c}^{\mu}\cos[\mu(\varphi-\varphi_{0})] \label{9}%
\end{equation}

Using atomic unit $\frac{me_{s}^{2}}{\hbar^{2}}$ with $e_{s}=e^{2}%
/(4\pi\varepsilon_{0})$, which is the dimension of length, we set $\sqrt
{\frac{2m\gamma_{\upsilon}e}{\hbar^{2}}}=1$ in the numerical evaluation. The
classical orbits with the initial\textbf{\ }angle setting to zero $\varphi
_{0}=0$ are shown in Figs. 1-5 (solid green curves), which depend on the
initial angular momentum $\xi_{k}$ only for given scalar potential.\textbf{
}In general, we have closed orbits (Figs. 1-4) for $\mu>0$ $(\upsilon>2)$,
while open trajectories for $\mu<-2$ seen from Fig.5. From Eq.(9) it is
obvious that the rotation symmetry of classical orbits depends on the
power-index $\mu$ such that the orbits are invariant under a rotation-angle
$\frac{2\pi}{|\mu|}$, and $2\pi$-symmetry holds only for $|\mu|=1$. We
demonstrate in this paper that the angular momentum quantization can be
determined only by the rotation symmetry of classical orbits based on the
requirement of quantum-classical correspondence. The rotation angles of
symmetry for closed orbits in Figs.1-4 are 2$\pi$, $6\pi$, $\frac{4\pi}{5}$,
and $\frac{2\pi}{5}$ respectively.

\section{ Angular momentum quantization}

In the polar coordinates $(r,\varphi)$, the zero-energy Schr$\ddot{o}$dinger
equation
\begin{equation}
\frac{-\hbar^{2}}{2m}[\frac{\partial^{2}}{\partial r^{2}}+\frac{1}{r}%
\frac{\partial}{\partial r}+\frac{1}{r^{2}}(\frac{\partial}{\partial\varphi
}-i\alpha)^{2}]\psi+eA_{0}\psi=0
\end{equation}
becomes%

\[
(\frac{\partial^{2}}{\partial r^{2}}+\frac{1}{r}\frac{\partial}{\partial
r}-\frac{\lambda^{2}}{r^{2}})R(r)+\frac{2me}{\hbar^{2}}\frac{\gamma_{\upsilon
}}{r^{\upsilon}}R(r)=0,
\]%
\[
(\frac{\partial}{\partial\varphi}-i\alpha)^{2}\Theta(\varphi)=-\lambda
^{2}\Theta(\varphi).
\]
by the separation of variables $\psi=R(r)\Theta(\varphi)$. The eigenvalue
solution of angular part Eq. (10) is
\begin{equation}
\Theta_{l^{c}}(\varphi)=N_{\varphi}e^{il_{c}\varphi}.
\end{equation}
with $N_{\varphi}$ being the normalization constant and the CAM eigenvalue
$l_{c}$ to be determined. For the usual requirement of 2$\pi$-periodic
boundary-condition, $\Theta(\varphi)=\Theta(\varphi+2\pi)$, one can obtain the
integer angular momentum quantization and the normalization constant
$N_{\varphi}=\frac{1}{\sqrt{2\pi}}$. However the 2$\pi$-periodic
boundary-condition is not justified in the $2D$-space, although the potential
$A_{0}(r)=-\frac{\gamma_{\upsilon}}{r^{2\mu+2}}$ is symmetric under rotation.
It has been demonstrated that a macroscopic quantum state, here SU(2) coherent
state \cite{Makowski1} can be constructed with probability density of wave
functions well localized on the classical orbits for the quantum-classical
correspondence, which \ results in a special boundary condition of the angular
momentum eigenstates such that rotational period of wave function is not
necessarily 2$\pi$ but should be the same as that of classical orbits
$\Theta(\varphi)=\Theta(\varphi+\frac{2\pi}{\left\vert \mu\right\vert }%
)$\cite{Makowski1}. Thus the CAM eigenvalue is no longer integer but should be
set as
\begin{equation}
l_{n}^{c}=n\left\vert \mu\right\vert , \label{12}%
\end{equation}
where $n$ is a integer. The angular momentum now is quantized with an
eigenvalue-space%
\begin{equation}
\Delta l=|\mu|
\end{equation}
The integer-quantization is only a special case of $|\mu|=1$. The
normalization constant becomes%
\begin{equation}
N_{\varphi}=\sqrt{\frac{\left\vert \mu\right\vert }{2\pi}}%
\end{equation}
The radial equation is%
\begin{equation}
(\frac{\partial^{2}}{\partial r^{2}}+\frac{1}{r}\frac{\partial}{\partial
r}-\frac{(l_{n}^{c}-\alpha)^{2}}{r^{2}})R(r)+\frac{2me}{\hbar^{2}}\frac
{\gamma_{\upsilon}}{r^{\upsilon}}R(r)=0,
\end{equation}
In the choice of Eq. (12) for $l_{c}$, the $n$-th KAM eigenvalue is shifted by
the flux number that
\begin{equation}
l_{n}^{k}=l_{n}^{c}-\alpha=n\left\vert \mu\right\vert -\alpha\label{16}%
\end{equation}
indicating the dynamic effect of the gauge potential in contradiction with the
classical solution in which gauge field does not apply a torque on the
particle. We do have the other choice of $l_{c}$ such that%
\begin{equation}
l_{n}^{c}=n\left\vert \mu\right\vert +\alpha\label{17}%
\end{equation}
while the KAM eigenvalue
\begin{equation}
l_{n}^{k}=n\left\vert \mu\right\vert \label{18}%
\end{equation}
does not depend on the flux in consistence with classical solution. The
common-phase $e^{i\alpha\varphi}$ factor called the topological phase
\cite{Liang1,Liang2} does not change the angular momentum quantization Eq.(13)
nor the normalization constant Eq.(14) since whole eigenfunctions have the
same additional angular phase. It is the main goal of the present paper that
only the choice of CAM Eq.(17) gives rise to the exact quantum-classical
correspondence. \ \textbf{\ \ \ }

\subsection{Topological phase of the gauge field and exact quantum-classical
correspondence}

We demonstrate in this section that only the CAM eigenvalues Eq.(17) consist
with the quantum-classical correspondence, such that the probability densities
of coherent wave functions are well localized on classical orbits. Introducing
the dimensionless radius $\chi=r/\tilde{a}_{q}$ and $y=1/\chi$ , where
$\tilde{a}_{q}$ is a quantity with dimension of length, we obtain
\begin{equation}
(y^{2}\frac{d^{2}}{dy^{2}}+y\frac{d}{dy}-(l_{n}^{^{k}})^{2}+B^{2}%
y^{\upsilon-2})R_{\tau}(y)=0 \label{19}%
\end{equation}
where the parameter $B$ is defined by%
\[
B^{2}\equiv\frac{2me\gamma_{\upsilon}}{\hbar^{2}\tilde{a}_{q}^{\upsilon-2}}%
\]
and the corresponding KAM eigenvalues $l_{n}^{k}$ given by Eq.(18) do not
depend on the flux number. Square-integrable solutions of Eq. (19) are found
in terms of Bessel functions of the first kind
\cite{Makowski7,Daboul1,Daboul3,Watson}%
\begin{equation}
R_{l_{n}^{k}}(y)=N_{l_{n}^{k}}J_{\frac{l_{n}^{k}}{\left\vert \mu\right\vert }%
}(\frac{1}{\left\vert \mu\right\vert r^{\mu}}) \label{20}%
\end{equation}
with the normalization constant given by
\begin{equation}
N_{l_{n}^{k}}=\sqrt{2\sqrt{\pi}\left\vert \mu\right\vert ^{2/\mu}\frac
{\Gamma(1+1/\mu)\Gamma\lbrack1+l_{n}^{k}/\left\vert \mu\right\vert +1/\mu
]}{\Gamma(1/2+1/\mu)\Gamma\lbrack l_{n}^{k}/\left\vert \mu\right\vert -1/\mu
]}}%
\end{equation}
where we have set the parameter $\sqrt{\frac{2m\gamma_{\upsilon}e}{\hbar^{2}}%
}=1,$ and thus $B\tilde{a}_{q}^{\mu}=1,\quad\tilde{a}_{q}^{2}B^{^{2/\mu}}=1$
according to the definition of $B$. If the following conditions are satisfied:%
\begin{equation}
\operatorname{Re}(\frac{2l_{n}^{k}}{\left\vert \mu\right\vert }%
+1)>\operatorname{Re}(\frac{2}{\mu}+1)>0, \label{22}%
\end{equation}
\textbf{ }we have bound states corresponding to the classical closed-orbits as
demonstrated in Ref \cite{Daboul2}. For the zero-energy $E=0$ there exist
classical solutions of closed-orbits with any non-zero angular momentum when
$\upsilon>2$, while the condition of bound states in quantum mechanics is
$l_{n}^{k}>1$\textbf{. }It has been shown that\textbf{ }\cite{Daboul3}\textbf{
}the normalizable quantum solutions can be classified to two classes: (1)
bound states ( $l_{n}^{k}>1$) for $\mu>0$ $(\upsilon>2)$ corresponding to the
classical closed-orbits (Figs. 1-4, solid green curves) (2) scattering states
($l_{k}\geqslant0)$ for $\mu<-2$ $(\upsilon<-2),$which are normalizable,
corresponding to classical open-orbits (Fig. 5, solid green curves). At the
region $-2\leq\upsilon\leq2$ the solutions of wave functions are not
square-integrable \cite{Daboul1,Daboul3}, which we do not discuss in the
present paper. The complete eigenfunctions of the Schr$\ddot{o}$dinger
equation\ can be written in the explicit form as
\cite{Makowski7,Daboul1,Daboul3,Watson}%
\begin{equation}
\psi_{\mu,l_{n}^{k}}(r,\varphi)=N_{\varphi}e^{i(l_{n}^{k}+\alpha)\varphi
}N_{l_{n}^{k}}J_{\frac{l_{n}^{k}}{\left\vert \mu\right\vert }}(\frac
{1}{\left\vert \mu\right\vert r^{\mu}}). \label{23}%
\end{equation}
\ Following Refs.\cite{Makowski1,Makowski2} we construct\textbf{ }the
stationary SU(2) coherent-state for the central-scalar-potentials in the
standard way as \cite{J.R}%
\begin{align}
\Psi_{\mu,N}(r,\varphi)  &  =\frac{1}{(2)^{N/2}}\label{24}\\
&  {\displaystyle}{\sum\limits_{n=0}^{N}}{\binom{N}{n}}^{1/2}N_{\varphi
}e^{i(l_{n}^{k}+c+\alpha)\varphi}N_{l_{n}^{k}+c}J_{\frac{l_{n}^{k}%
+c}{\left\vert \mu\right\vert }}(\frac{1}{\left\vert \mu\right\vert r^{\mu}%
})\nonumber
\end{align}
with $l_{n}^{c}$ and $l_{n}^{k}$ given in Eqs.(17), (18) respectively. An
additional angular momentum number
\begin{equation}
c=n_{c}\left\vert \mu\right\vert \label{25}%
\end{equation}
with $n_{c}$ being a integer-parameter is introduced to adjust the probability
density, which does not change the rotational symmetry of wave-function
density but the spatial expansion of it. In the following section we will see
that the parameter $n_{c}$ is not arbitrary but is related to the classical
KAM $\pounds ^{k}$ such that the expectation value of KAM operator coincides
with the classical value.\textbf{ }The wave functions are normalized to unity
in the angular range of $\varphi$ being from $-\pi$ to $\pi$\textbf{. }From
Eq. (24) we can see that the common topological phase-factor resulted by the
gauge potential does not affect the probability density of wave functions in
agreement with the classical assumption of no torque\textbf{ }applied on the
charged particle. In Figs. 1-5 the practical values of parameters are given by
$\mu=1,\alpha=6\frac{1}{2},n_{c}=8$, $\mu=\frac{1}{3},\alpha=\frac{5}{4},$
$n_{c}=9$, $\mu=\frac{5}{2},\alpha=31\frac{1}{3},$ $n_{c}=14$, $\mu
=5,\alpha=55\frac{3}{5},n_{c}=12$, and $\mu=-\frac{7}{3},\alpha=42\frac{7}%
{10},$ $n_{c}=19$ respectively. In all figures the total number of
eigenfunctions is $N=30$. The density $|\Psi_{\mu,N}(r,\varphi)|^{2}$ of
coherent-state wave function possesses the same rotation symmetry of the
classical orbits. Figs. 1-5 (a) show that the probability densities of
coherent wave functions are well localized on the classical orbits in each
case indicating the exact quantum-classical correspondence. We thus obtain the
angular-momentum quantizations with the level-spaces $1,\frac{1}{3},\frac
{5}{2},5,$ and $\frac{7}{3}$ (of $\hbar$) respectively. The CAM spectrum is
shifted by the corresponding flux number $\alpha$.

\textbf{ }

\begin{figure}[ptb]
%begin{center}
\includegraphics[height=3.8in,width=2.6in]{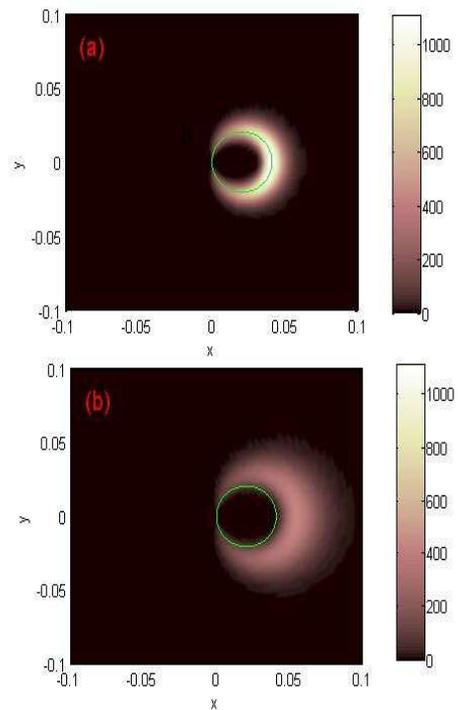}
%\end{center}
\caption{(Colour online) Probability density image $\left\vert
\Psi_{\mu ,N}(r,\varphi)\right\vert ^{2}$ and the closed
classical-orbit (solid green curve) for
$\mu=1(\upsilon=4),\alpha=6\frac{1}{2},\xi_{k}=23$. (a) shift of
CAM; (b) shift of KAM.}%
\end{figure}
\begin{figure}[ptbptb]
%\begin{center}
\includegraphics[height=4.0in,width=2.8in]{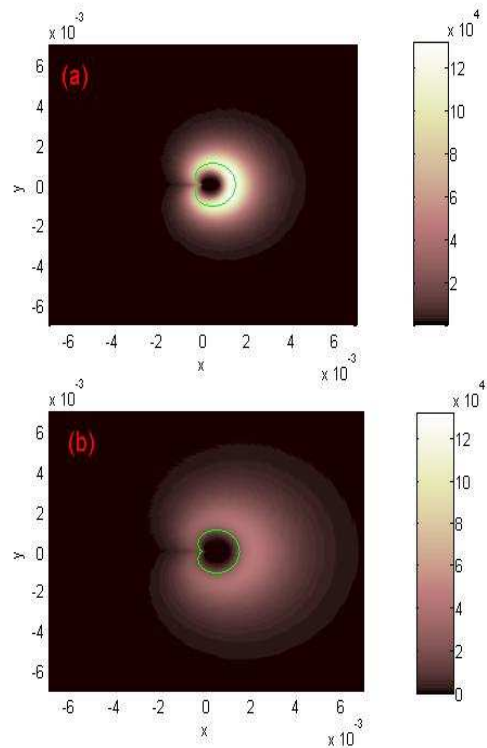}
%end{center}
\caption{(Colour online)
$\mu=\frac{1}{3}(\upsilon=\frac{8}{3}),\alpha
=\frac{5}{4},\xi_{k}=8$.}%
\end{figure}\
\begin{figure}[ptbptbptb]
\begin{center}
\includegraphics[height=3.8in,width=2.6in]{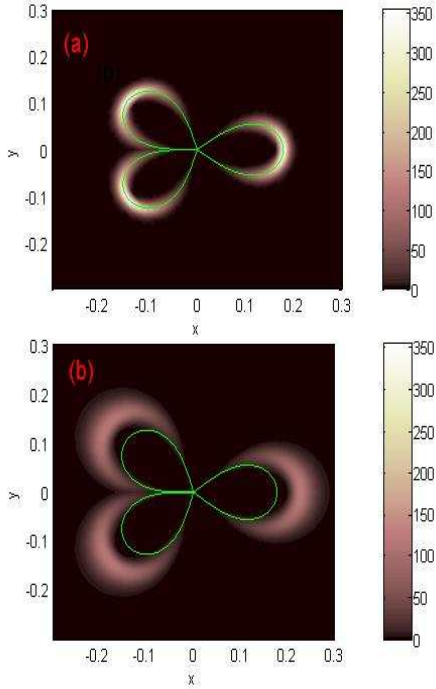}
\end{center}
\caption{(Colour online) $\mu=\frac{5}{2}(\upsilon=7),\alpha=31\frac{1}{3}%
,\xi_{k}=72.5$.}%
\end{figure}
\begin{figure}[ptbptbptbptb]
%\begin{center}
\includegraphics[height=3.8in,width=2.6in]{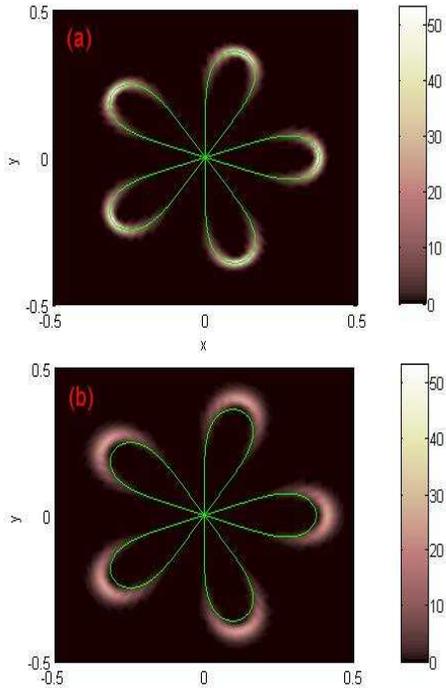}
%\end{center}
\caption{(Colour online) $\mu=5(\upsilon=12),\alpha=55\frac{3}{5},\xi_{k}%
=135$.}%
\end{figure}

\begin{figure}[ptb]
%\begin{center}
\includegraphics[height=3.8in,width=2.6in]{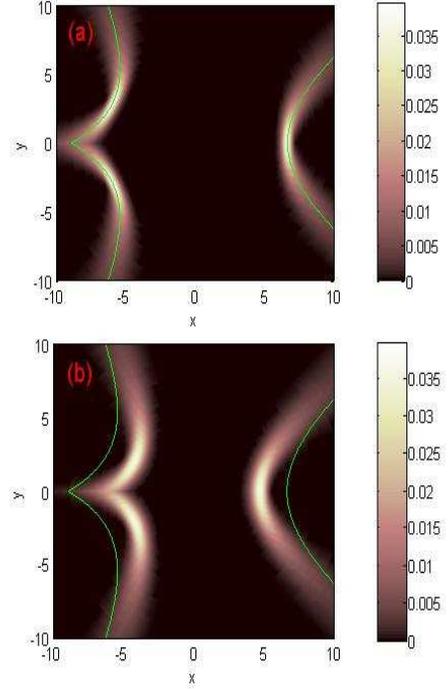}
%\end{center}
\caption{(Colour online) Probability density image $\left\vert
\Psi_{\mu ,N}(r,\varphi)\right\vert ^{2}$ and the open
classical-orbit (solid green
curve) for $\mu=-\frac{7}{3}(\upsilon=-\frac{8}{3}),\alpha=42\frac{7}{10}%
,\xi_{k}=\frac{238}{3}\ .$(a) shift of CAM (b) shift of KAM.}%
\end{figure}

Besides the AB interference \cite{Liang1} the fractional spins have many
important physical applications, for example, two-vortex system in a
superfluid film \cite{Leinaas}, necklace-ring \cite{He,Desyatnikov},
chiral-wave superconductor\cite{Goryo} or quantum billiard\cite{Gongora}
inside a boundary defined by the wedge-shaped section of a circle.

\emph{Shift of the KAM eigenvalues by the gauge potential and break down of
the correspondence: }We, of course, can have the other choice of CAM
eigenvalues i.e. Eq.(12), which does not depend on the flux and as a
consequence the spectrum of KAM is shifted by the gauge potential as shown in
Eq.(16). In this case we do not have the topological phase of the wave
functions but have the flux-dependent KAM eigenvalues instead. The
quantization of angular momentum, namely the level-space of eigenvalues, is
not changed by the gauge potential. Replacing the KAM eigenvalues $l_{n}^{k}$
in Eqs.(23,24) by Eq.(16) the corresponding probability densities of coherent
states are depicted in Figs.1-5 (b), from which we see that probability
densities are no longer localized no the classical orbits indicating
additional torques applied on the particle in contradiction with the classical assumption.

\emph{Expectation values of angular momentum}: We now calculate the
expectation values of KAM operator%
\[
\mathbf{\ }\hat{\pounds }^{k}=-i\hbar\lbrack\frac{\partial}{\partial\varphi
}-i\alpha]
\]
in the SU(2) coherent-state to find the explicit relation between the
adjusting-parameter $c$ and the initial classical-KAM $\pounds ^{k}=\xi
_{k}\hbar$ in order to confirm further the quantum-classical correspondence.
The average of KAM operator\textbf{ }in the SU(2) coherent state Eq.(24) with
the topological phase is evaluated as%

\begin{equation}
\left\langle \hat{\pounds }^{k}\right\rangle =\left\langle \Psi_{\mu
,N}\left\vert \hat{\pounds }^{k}\right\vert \Psi_{\mu,N}\right\rangle
=\hbar\left\vert \mu\right\vert [n_{c}+\frac{N}{2}]
\end{equation}
Substituting the corresponding parameter values of $\mu$, $n_{c}$, and $N$
into the above formula one can see the exact agreement with the initial-values
of classical KAM such that%
\begin{equation}
\left\langle \hat{\pounds }^{k}\right\rangle =\xi_{k}%
\end{equation}
for $\xi_{k}$= $23,8,72.5,135,\frac{238}{3}$respectively. However, for the CAM
eigenvalue-choice of Eq.(12), namely the shift of KAM eigenvalue by the gauge
potential\textbf{, }the average of KAM operator becomes%
\begin{equation}
\left\langle \hat{\pounds }^{k}\right\rangle =\hbar\lbrack(n_{c}+\frac{N}%
{2})\left\vert \mu\right\vert -\alpha]
\end{equation}
\newline which, of course, disagrees with classical values.

\section{Conclusion}

In summary, the NIQAM in two-dimensional space is demonstrated in terms of
quantum-classical correspondence with an exactly solvable model, such that the
probability clouds of macroscopic quantum states, here the SU(2) coherent wave
functions, are well localized on the classical orbits. The eigenfunctions of
angular momentum have to possess the same rotational symmetry of classical
orbits, in which the rotation-period can be greater or less than $2\pi$
depending on the power-index $\mu$ of the central potential only. As a
consequence the level-space of angular momentum spectrum is less or greater
than $\hbar$ and the integer-quantization (level-space $\hbar$) is possible
only if $\mu=1$. The gauge potential of AB-flux does not affect the angular
momentum quantization but can shift the spectrum of angular momentum by a
common value. By explicit calculations it is shown that the quantum-classical
correspondence results in the unambiguous determination of CAM eigenvalues
with a common topological-phase in the wave functions, the probability
densities of which coincide with the classical orbits for any power-index of
potentials. The quantum mechanical model of anyon proposed by Wilczek
\cite{Wilczek1,Wilczek2} and latter clarified as the fractional CAM in
Ref.\cite{Liang1} becomes a special case of the present model with $\mu=1$.

\section{Acknowledgment}

This work was supported by National Nature Science Foundation of China (Grant No.10775091).

\end{document}